\documentclass[sigconf]{acmart}

\AtBeginDocument{%
  \providecommand\BibTeX{{%
    \normalfont B\kern-0.5em{\scshape i\kern-0.25em b}\kern-0.8em\TeX}}}

\acmBooktitle{}

\usepackage{graphicx}
\usepackage{subfigure}
\usepackage{soul}
\usepackage{ulem}
\usepackage{xcolor}

\newcommand{\cataractbot}{\textit{CataractBot}}
\newcommand{\anonymousHospital}{Sankara Eye Hospital}
\newcommand{\link}[1]{\textcolor{blue}{#1}}

\usepackage{setspace} 
\usepackage{etoolbox}

\newcommand{\bheading}[1]{\vspace*{.5em}\noindent{\textbf{#1.}}}

\begin{document}

\title[Large-Scale Deployment of an LLM-Powered Expert-in-the-Loop Healthcare Chatbot]{Learnings from a Large-Scale Deployment of an LLM-Powered Expert-in-the-Loop Healthcare Chatbot}

\author{Bhuvan Sachdeva}
\orcid{0009-0002-1946-684X}
\authornotemark[1]
\affiliation{%
  \institution{Microsoft Research}
  \city{Bangalore}
  \country{India}}
\email{b-bsachdeva@microsoft.com}

\author{Pragnya Ramjee}
\authornote{Both authors contributed equally to this research.}
\orcid{0000-0003-0061-2624}
\affiliation{%
  \institution{Microsoft Research}
  \city{Bangalore}
  \country{India}
}
\email{t-pramjee@microsoft.com}

\author{Geeta Fulari}
\affiliation{%
  \institution{Sankara Eye Foundation India}
  \city{Bangalore}
  \country{India}}
\email{quality@sankaraeye.com}

\author{Kaushik Murali}
\affiliation{%
  \institution{Sankara Eye Foundation India}
  \city{Bangalore}
  \country{India}}
\email{kaushik@sankaraeye.com}

\author{Mohit Jain}
\orcid{0000-0002-7106-164X}
\affiliation{%
  \institution{Microsoft Research}
  \city{Bangalore}
  \country{India}}
\email{mohja@microsoft.com}

\renewcommand{\shortauthors}{Sachdeva et al.}

\begin{abstract}
  Large Language Models (LLMs) are widely used in healthcare, but limitations like hallucinations, incomplete information, and bias hinder their reliability. To address these, researchers released the Build Your Own expert Bot (BYOeB) platform, enabling developers to create LLM-powered chatbots with integrated expert verification. CataractBot, its first implementation, provides expert-verified responses to cataract surgery questions. A pilot evaluation showed its potential; however the study had a small sample size and was primarily qualitative. In this work, we conducted a large-scale 24-week deployment of CataractBot involving 318 patients and attendants who sent 1,992 messages, with 91.71\% of responses verified by seven experts. Analysis of interaction logs revealed that medical questions significantly outnumbered logistical ones, hallucinations were negligible, and experts rated 84.52\% of medical answers as accurate. As the knowledge base expanded with expert corrections, system performance improved by 19.02\%, reducing expert workload. These insights guide the design of future LLM-powered chatbots.

\end{abstract}




\maketitle

\section{Introduction}
Since the advent of large language models (LLMs) in late 2022, these models have experienced rapid and widespread adoption.
OpenAI's ChatGPT, for instance, reached 1 million users within just five days of its launch, and within two months, it surpassed 100 million users, making it the fastest-growing consumer application in history~\cite{forbes2023chatgpt}.
The unprecedented success of LLMs is primarily attributed to their ease of use, ability to understand natural language, and extensive knowledge base, which enables them to answer a broad range of questions effectively~\cite{kabir2024chatptstackoverflow,yang2023talk2care,park2024chatgptinhighereducation}.
These capabilities have led to a widespread adoption of LLMs in end-user applications and chatbots, such as software development~\cite{TheEconomicGithubCopilot}, healthcare~\cite{mayoclinic}, and education~\cite{khanmigo, duolingomax}.
Despite their utility, LLMs have limitations, including hallucinating, providing incomplete or outdated information, struggling with complex questions, being inconsistent, and exhibiting obscured reasoning and bias~\cite{denecke2024potentialofllms, au2023ai, fogel2018artificial, iliashenko2019opportunities, kocaballi2020envisioning}.
Such issues are especially concerning in healthcare settings, where accuracy and trustworthiness are critical.
Hence, leading organizations like OpenAI and Anthropic have implemented usage policies stating that `\textit{tailored medical or health advice cannot be provided without review by a qualified professional}'~\cite{openai_usage_policies}.

To address these concerns and comply with these policies, researchers introduced an open-source platform `Build Your Own expert Bot' (BYOeB)~\cite{byoeb_github}, enabling developers to create expert-in-the-loop LLM-powered chatbots.
The platform's first application was \cataractbot{}~\cite{ramjee2024cataractbot}, a WhatsApp-based chatbot designed to help patients and their attendants with queries related to cataract surgery.
The bot provided verified answers for medical questions from doctors and for logistical questions from patient coordinators.
These expert-provided answers were used to refine the custom knowledge base to minimize future expert intervention.
To explore \cataractbot{}'s usage, impact, and adoption, \citet{ramjee2024cataractbot} conducted a preliminary 8-week study involving 55 users at an eye hospital in India.
The study found that patients and attendants appreciated that \cataractbot{} reduced their hesitation to ask questions, which was often due to power dynamics. 
Their trust in the system was established through the verification performed by doctors and coordinators.
Experts, on the other hand, commended the bot for acting as a mediator, providing a layer of privacy between them and the patients.
However, this study was primarily qualitative and relied extensively on interview data, which limits the generalizability of its findings.
Additionally, the presence of researchers during interviews has been shown to introduce bias amongst participants and influence their responses~\cite{thies-bias-chi}.
In contrast, our in-the-wild longitudinal study with a larger sample size offers the potential to generate more robust, quantitative, real-world insights, providing a stronger foundation for understanding the bot's true effectiveness in practical settings.

In this work, we extend the previous research by conducting a large-scale, in-the-wild study of \cataractbot{} in a clinical setting.
Over a six-month period, 318 patients and attendants interacted with the system, sending 1,992 messages that were verified by 5 doctors and 2 patient coordinators.
This paper presents findings from both quantitative and qualitative analyses of the interaction logs, addressing the following research questions:
\textbf{RQ1}: How did the system perform?
\textbf{RQ2}: How did end-users (patients and attendants) use the system?
\textbf{RQ3}: How did experts (doctors and patient coordinators) engage with the system?
Our analysis revealed that patients and attendants mainly used the bot to address their medical rather than logistical questions, with activity peaking on the day before surgery.
Experts corrected 18.06\% of the LLM-generated answers, by adding new information in 76.08\% of those corrections.
This improved system performance, with the proportion of answers marked as `accurate and complete' increasing from 65.60\% in the first four weeks to 84.62\% in the last four weeks of deployment.
However, experts frequently overlooked patient-specific questions, as their corrections would not update the knowledge base or reduce their workload.
As the knowledge base grew, we noticed repeated content and conflicting recommendations from different doctors.
We discuss design considerations to tackle such challenges and provide insights for designing LLM-powered expert-in-the-loop chatbots.


\section{Related Work}

HCI and healthcare researchers emphasize the importance of addressing patients' informational needs. 
Patients and their attendants seek to understand their health conditions, diagnoses, treatment options, potential risks, and preventive measures~\cite{info-needs-review}.
This understanding has been found to improve their treatment adherence and overall satisfaction with healthcare services~\cite{informing-patients}.
While patients typically prefer to rely on the advice of healthcare professionals, in-person consultations places a significant burden on healthcare systems and are often costly in terms of both time and money for patients~\cite{akerkar2004doctorpatientrelationship,bodenheimer2013primary}.
Moreover, the increasing pressure on doctors to accommodate more patients has led to reduced time per patient, negatively affecting communication and information sharing. 
Studies have shown that doctors often underestimate patients' information needs and overestimate how much information they provide~\cite{Tongue2005CommunicationSF, info-need-before-consult}.
These consultations also do not always fully address patients' questions due to power imbalances between doctors and patients~\cite{fochsen2006imbalance} and the challenge of information overload during these brief interactions~\cite{preanesthetic-info-ferre-2020, info-need-before-consult}.

Researchers have explored various technological solutions to bridge the communication gaps between patients and their healthcare providers, including phone calls~\cite{ivatury2009healthhotlines}, IVRs~\cite{joshi2014ivrsforhivpositive}, SMS~\cite{perrier2015kenyapregnantwomensms}, emails~\cite{patt2003emailfordoctors,makarem2016email}, social media chat groups~\cite{ding-wechat-chi2020,grainger2017facebookdiscussion}, and video calls~\cite{ding2020askingdoctorsonline,gordon2020telehealthnotfeelingpart}.
For instance, \citet{joshi2014ivrsforhivpositive} developed an IVR-based system in India to support people living with AIDS, providing health tips and a channel to report symptoms. 
However, IVRs are often limited by their inflexible, sequential interactions and inability to save information for later referencing~\cite{nngroup2023phonetree}.
\citet{ding-wechat-chi2020} studied nurse-facilitated patient groups on WeChat (a messaging platform) in China, where nurses extended care and promoted peer support.
This added to nurses' workloads, requiring them to sift through chat histories, locate actionable messages, and respond individually, leading to delays and missed messages.
The presence of peers in these groups prompted some patients to initiate private chats with nurses for sensitive issues, further increasing their workload.

Chatbots address many of these limitations by enabling real-time, natural language interactions with minimal learning curves, offering a personalized and flexible experience~\cite{wutz2023factors}.
They have been applied in various healthcare settings, including appointment scheduling~\cite{dammavalam2022chatbotforhospitalmanagement}, providing information~\cite{bickmore2009virtualnurseagents,yadav-feeding-2019}, and supporting mental health~\cite{lee2020chatbotformentalhealthdisclosure,inkster2018wysarealworld}.
For instance, \citet{bickmore2009virtualnurseagents} created an animated virtual nurse interface to educate patients during discharge, and \citet{yadav-feeding-2019} explored the potential of a chatbot for breastfeeding education.
As these chatbots were rule-based, they were constrained by predefined conversation flows.
The emergence of LLMs, with their superior natural language understanding and ability to engage in open-ended conversations with access to a vast body of information, has renewed interest in the application of chatbots within healthcare environments~\cite{jo2023understanding,yang2023talk2care,wei2024selfreporteddatallms}.
Researchers have developed LLM-powered systems to help healthcare providers in the clinical decision-making process~\cite{rajashekar2024llmpoweredecisionmaking}, analyse medical images~\cite{yildirim2024radiologydetection}, and facilitate patient-provider communication~\cite{yang2023talk2care,jo2023understanding,ramjee2024cataractbot}.
For instance, \citet{rajashekar2024humanalgorithmic} developed a decision-support system for physicians, and found the LLM interface enhanced ease of use, and \citet{yang2023talk2care} developed an LLM chatbot to collect health information from older adults and present to their healthcare providers.
However, these studies are limited to research settings with small sample sizes (13 to 55 participants)~\cite{rajashekar2024humanalgorithmic,yildirim2024radiologydetection,yang2023talk2care,jo2023understanding,ramjee2024cataractbot} and rely on qualitative analysis, which can introduce participant bias~\cite{thies-bias-chi}.

Recently, studies have deployed healthcare chatbots in real-world settings with larger sample sizes.
For instance, \citet{beatty2022wysatherapeuticalliance} conducted a mixed-methods study with 1,205 participants using the commercial mental health chatbot Wysa\footnote{Wysa: \link{\url{https://www.wysa.com}}}. 
Their research found positive changes in therapeutic alliance scores based on in-app questionnaires and conversation transcript analysis.
\citet{jo2023understanding} highlighted the effectiveness of their LLM-powered chatbot in supporting socially isolated individuals through check-up phone calls, deployed with 301 participants in Seoul.
Outside mental health, \citet{fan2021chatbotrealworld} analyzed interaction logs of a self-diagnostic rule-based chatbot in China used by 16,519 users over 6 months. Similarly, \citet{powering-ai-chatbot-xiao-2023} developed an AI-powered chatbot, crowdsourcing expert knowledge to address the public's information needs during the COVID-19 pandemic, handling 1,252 conversations over 6 months.
However, many of these systems lack expert oversight, raising concerns about AI reliability in healthcare, including issues of reasoning transparency, inconsistency, errors, hallucinations, and bias~\cite{au2023ai}.
To mitigate these risks, recently, \citet{ramjee2024cataractbot} proposed a human-AI collaboration approach with \cataractbot{}, where LLM-generated answers to healthcare queries are verified by an expert.
They conducted a small-scale pilot deployment with 55 users.
In our work, we conduct a large-scale, in-the-wild deployment study of \cataractbot{} with 321 patients and attendants, examining its use and performance in a clinical setting.

\section{Deployment Details}

\subsection{Method}
The \cataractbot{} system was deployed at \anonymousHospital{} in Bangalore, India, which caters to patients from diverse linguistic, educational, and technical backgrounds.
As per hospital protocol, once a patient opts for cataract surgery based on a doctor's recommendation, the patient and their attendant meet with a patient coordinator.
The coordinator schedules the surgery and provides guidance on pre- and post-operative measures.
At the end of this interaction, the coordinator assessed the patient's eligibility for \cataractbot{} based on these criteria: aged 18 or above, fluent in one of the five languages supported by \cataractbot{} (English, Hindi, Kannada, Tamil, and Telugu), and scheduled to undergo surgery with one of the 5 participating operating doctors.
If these conditions were met, the coordinator suggested the use of \cataractbot{} to address their surgery-related queries.
Upon obtaining consent, the coordinator filled a web-based onboarding form, which included preferred language, WhatsApp numbers, surgery date, and demographic details.
After form submission, participants received `onboarding messages' from \cataractbot{}.
Participants were instructed to ask a trial question and the coordinator briefly explained the bot's icons and expert verification system. 
Post-onboarding, participants received reminder messages at 4pm on five specific days---the day after onboarding, the day before surgery, the surgery day, the day after surgery, and five days post-surgery---reminding them of \cataractbot{}'s availability for surgery-related questions.

\subsection{\cataractbot{} System}
All features of \cataractbot{} have been previously described in detail~\cite{ramjee2024cataractbot, byoeb_github}.
We provide a summary here.

\cataractbot{} supports three interaction modalities: text, speech, and tap.
For every voice message, \cataractbot{} provides both a text and an audio response.
For tap-based interactions, the bot suggests three questions with the welcome message and three related questions after each response, enabling users to tap and continue the conversation.
Upon receiving a message, \cataractbot{} classifies it as a medical question (e.g., post-surgery care), logistical question (e.g., scheduling or insurance), or small talk, and responds in real-time.
For medical and logistical questions, the bot strictly employs the knowledge base curated by the hospital to generate an appropriate response, which is marked as unverified.
When the custom knowledge base lacks an answer, the bot responds with a template ``\textit{I don't know}'' response.
For small talk messages (such as ``\textit{Hello}'' or ``\textit{Thank you for the info}''), \cataractbot{} provides corresponding small talk responses.

For medical questions, the operating doctor receives a message, including the question asked, the bot's response, and patient's demographics.
The doctor is asked, ``\textit{Is the answer accurate and complete?}'' with three response options: `\textit{Yes}', `\textit{No}', and `\textit{Send to Patient Coordinator}'.
Selecting `\textit{Yes}' notifies the patient that the answer has been verified.
Selecting `\textit{No}'
alerts the patient to await a corrected response.
The doctor is asked to provide a correction in free-form text, which \cataractbot{} automatically combines with the bot's initial answer to create a new response, delivered to the patient (Figure \ref{fig:expertverifications}C,D).
If a question is misclassified, i.e., a logistical question is sent to the doctor, they can select `\textit{Send to Patient Coordinator}'.
Patient coordinators follow a similar workflow, verifying and correcting the bot's responses for logistical questions.

The \cataractbot{} system employs mechanisms to ensure timely verification.
If the operating doctor does not verify an answer within three hours, it is automatically also sent to the designated escalation doctor.
If neither doctor verifies the question within six hours, they both receive a reminder notification about the pending status.
Additionally, at 8 am, 12 pm, and 4 pm, a list of all questions pending for over six hours is sent to both doctors.
This workflow is mirrored for the patient coordinator and escalation patient coordinator.
To minimize experts' labor, expert-provided edits are used to update the knowledge base, increasing the likelihood of `Yes' responses from experts for similar questions.
A senior cataract surgeon, serving as the `knowledge base update expert',
reviews and selects expert-verified question-answer pairs for inclusion in the knowledge base. They also modify the answers as needed (Figure \ref{fig:expertverifications}E).

\subsection{Participants}
Although 550 people were onboarded on the bot, 318 (57.8\%) of them sent at least one message, which forms our participant set.
Among them, there were 154 patients and 164 attendants, comprising a total of 271 patient-attendant pairs.
A notable demographic difference was that attendants were generally younger, more fluent in English, and better educated compared to patients.
For details, please refer Table~\ref{tab:demography}.
Additionally, the study involved 5 doctors (4 as operating doctors, and 1 as escalation doctor and knowledge base update expert) and 2 patient coordinators (who alternated between coordinator and escalation coordinator roles).

\begin{table*}[]
\centering
\caption{Demography \& usage details of participants. (*63.52\% of participants provided this optional information; **Knowledge Base)}
\label{tab:demography}
\resizebox{\textwidth}{!}{%
\begin{tabular}{l|l|l}
\hline
                       & \textbf{Patients (n=154)}                                      & \textbf{Attendants (n=164)}                                    \\ \hline
\textbf{Age, Gender}   & 63.84$\pm$9.70 years, 47.40\% female                          & 37.88$\pm$10.69 years, 29.80\% female                          \\
\textbf{Education*}     & 22 $\leq$Grade 10, 18 Grade 12, 49 Bachelors, 10 Masters & 6 $\leq$Grade 10, 4 Grade 12, 45 Bachelors, 47 Masters, 1 PhD \\
\textbf{Language}      & 120 English, 11 Kannada, 10 Hindi, 7 Tamil, 6 Telugu & 148 English, 7 Kannada, 3 Hindi, 2 Tamil, 4 Telugu\\ \hline
\textbf{Message Type} & 653 medical, 272 logistical, 121 small-talk     & 615 medical, 257 logistical, 74 small-talk     \\
\textbf{Message Modality} & 618 text, 357 tap, 71 audio                             & 544 text, 379 tap, 23 audio                             \\
\textbf{Message Language} & 878 English, 39 Kannada, 46 Hindi, 59 Tamil, 24 Telugu        & 850 English, 23 Kannada, 18 Hindi, 15 Tamil, 40 Telugu        \\ \hline
\textbf{Expert Verification} & \multicolumn{2}{l}{1347 `\textit{Yes}', 319 `\textit{No}', 112 `\textit{Send to Doctor/Coordinator}', 131 no answer} \\ 
\textbf{KB** Update Expert} & \multicolumn{2}{l}{205 `\textit{Yes}', 85 `\textit{No}'; Among `\textit{Yes}': 104 edits} \\
\hline
\end{tabular}%
}
\end{table*}

\subsection{Data Analysis}
Our dataset consists of 1,992 messages sent by patients and attendants, verified by experts, and further approved by the knowledge base update expert.
For details, please refer Table~\ref{tab:demography}.
The data was analyzed quantitatively using descriptive statistics, t-tests, and (RM-)ANOVAs.
Additionally, one of the authors conducted a qualitative thematic analysis of 96.8\% of the data, by open-coding patients' questions, the bot's responses, and the edits performed by both the operating experts and the knowledge base update expert.
\section{Findings}
Here, we present key findings from the quantitative and qualitative analysis of the interaction logs.

\subsection{System Performance}

Below, we discuss \cataractbot{}'s performance across three LLM tasks and one AI-based language technology task.

\bheading{Response Generation}
An LLM was leveraged to generate responses to all queries posed by patients and attendants, based on a knowledge base curated by medical professionals.
Each generated response was verified by an expert.
Experts found most LLM-generated responses to be `accurate and complete' for medical questions (84.52\%) and logistical questions (69.46\%), and provided corrections for 187 (14.84\%) medical and 114 (28.07\%) logistical answers (discussed in Section~\ref{sec:finding-expert}).
On average, the bot responded in 9.27$\pm$4.90 seconds.
The bot provided an ``\textit{I don't know}'' response to 9.30\% of medical questions and 23.25\% of logistical questions.
This suggests that the higher number of ``\textit{I don't know}'' responses for logistical queries may account for the lower accuracy and completeness of those responses.
Further analysis revealed that a majority of these (37.50\%) ``\textit{I don't know}'' responses were due to gaps in the knowledge base, as seen in questions like ``\textit{What are some examples of light yoga postures I can do after surgery?}'',
and 29.38\% were due to patient-specific questions---e.g.,``\textit{What is the specific sugar level I should maintain before surgery?}''---which the bot could not answer without access to the patients' medical records.
Over the 24-week study, as the knowledge base was updated, the number of ``\textit{I don't know}'' responses decreased by 7.84\%  (Figure \ref{fig:charts_botperformance_querytypedistribution}A), and the number of LLM-generated answers that experts marked as `accurate and complete' increased by 19.02\%.

Manual analysis identified hallucinations in only 5 queries (0.25\%).
For instance, when a patient asked, ``\textit{Reports?}'', \cataractbot{} incorrectly responded, ``\textit{Your test reports have already been shared with the surgical team. They will take all necessary preventive measures before your surgery to avoid any risk factors.}''

\bheading{Message Classification}
Experts found the bot's classification of medical versus logistical queries to be inappropriate for only 6.23\% of questions.
Of these, they selected `\textit{Send to Patient Coordinator}' for 88.39\% and `\textit{Send to Doctor}' for the remainder.
This indicates that more logistical questions were misclassified as medical, rather than the reverse. 
Manual analysis of 60 random misclassified questions revealed the bot's classification to be correct in all cases, but doctors chose to delegate simple medical questions---e.g., ``\textit{What kind of light meal is recommended before surgery?}''---to coordinators.

\bheading{Expert Correction Incorporation}
To evaluate the bot's ability to incorporate expert corrections, we randomly selected 60 corrections 
performed by operating experts.
In all cases, the bot appropriately incorporated the expert's correction when generating a response for the patient.
This finding supports the design decision of~\cite{ramjee2024cataractbot} to not show the updated answer to experts, minimizing their workload.

\bheading{Transcription and Translation}
The \cataractbot{} system employed AI-based language technologies for transcribing audio questions and translating Indic language queries.
Limitations of these technologies resulted in 18.64\% of Indic languages queries and 26.74\% of audio-based queries receiving ``\textit{I don't know}'' responses.
Manual analysis revealed three common language-related errors in user input: using English despite choosing an Indic language (4.46\% of questions), using an Indic language in Latin script (0.41\%), and using an Indic language despite selecting English (0.30\%).
Despite these issues, \cataractbot{} correctly handled 77.00\% of these misformatted messages. 
In some cases, it even recognized the error.
For instance, when a patient asked about eye drops in Telugu but used the Latin script: ``\textit{Drops yenni ml veskovali?}'', \cataractbot{} replied, ``\textit{I'm sorry, but it seems like you're typing in another language. Could you please ask your question in English?}''
(Note: Raw audio messages were not stored due to privacy concerns, limiting further analysis of the audio modality.)

\subsection{End-User Interaction}
With respect to message distribution, participants asked significantly more medical questions (4.09$\pm$5.63 questions/participant) compared to logistical questions (1.70$\pm$2.24 questions/participant), with $t_{309}$=7.27, p<0.001.
This confirms that the bot was primarily used for addressing medical concerns.

\bheading{Modality}
A one-way RM-ANOVA of question modalities revealed a significant difference with $F_{2,618}$=66.68, p<0.001. 
Tukey's pairwise comparison showed that text (3.15$\pm$3.59) was the most commonly used modality, significantly more than taps (2.37$\pm$4.44) and speech (0.28$\pm$1.10) with p<0.01.
Taps were also preferred over speech (p<0.001).
The reluctance to use the speech modality may stem from concerns about the bot's limited transcription capabilities or privacy issues~\cite{ramjee2024cataractbot}.
Analysis of message logs revealed a typical pattern.
Most participants inquired about their surgery schedule the day before surgery.
(Note: As per the hospital protocol, attendants were supposed to receive a confirmation phone call on the day before surgery.)
Since the bot lacked access to personal details, it responded with ``\textit{I don't know}''.
Feeling uncertain, attendants often tapped on related questions in search of more information, even asking the same question multiple times.
For example, one attendant texted, ``\textit{When do I need to come to the hospital?}'' followed by nine taps, including ``\textit{What is the contact number for the inpatient department?}''.

\bheading{Demographics}
We found no significant difference in the number of questions asked by males and females, a positive finding given the existing gender disparity in accessing cataract surgery in India~\cite{prasad2020genderdifferencesincataract}.
However, there was a marginally significant difference based on education level: participants with a Bachelors degree or higher asked more questions (6.17$\pm$0.47 questions/participant) than those with 12th grade education or below (4.33$\pm$0.83), with $t_{197}$=1.92, p=0.056.
A two-way ANOVA between education level and question type ($F_{1,387}$=3.35, p=0.069), followed by Tukey's pairwise comparisons, showed a significant effect for medical questions (p<0.05), but no effect for logistical questions. 
This is expected, as medical questions typically involved more technical language (e.g., ``\textit{How is increased intraocular pressure diagnosed}''), whereas logistical questions were simpler (e.g., ``\textit{When is my surgery?}'').
Additionally, no significant differences were observed in the number of questions asked across different languages, supporting the decision to offer multilingual functionality. 
Our analysis revealed no significant interactions between modality choice and language (English vs. Indic), nor between modality and education level.
This is surprising, given that Indic keyboards are cumbersome to use~\cite{dalvi2015virtualkeyboardsIndian}, which might lead Indic users to prefer taps or audio modality, and that audio interactions are often recommended for low-literate users~\cite{farmchat-imwut18, findlater2009semiliterateaudiotext}.

\bheading{Thematic Analysis of Messages}
\begin{figure*}[t]
    \centering
    \includegraphics[width=.8\linewidth]{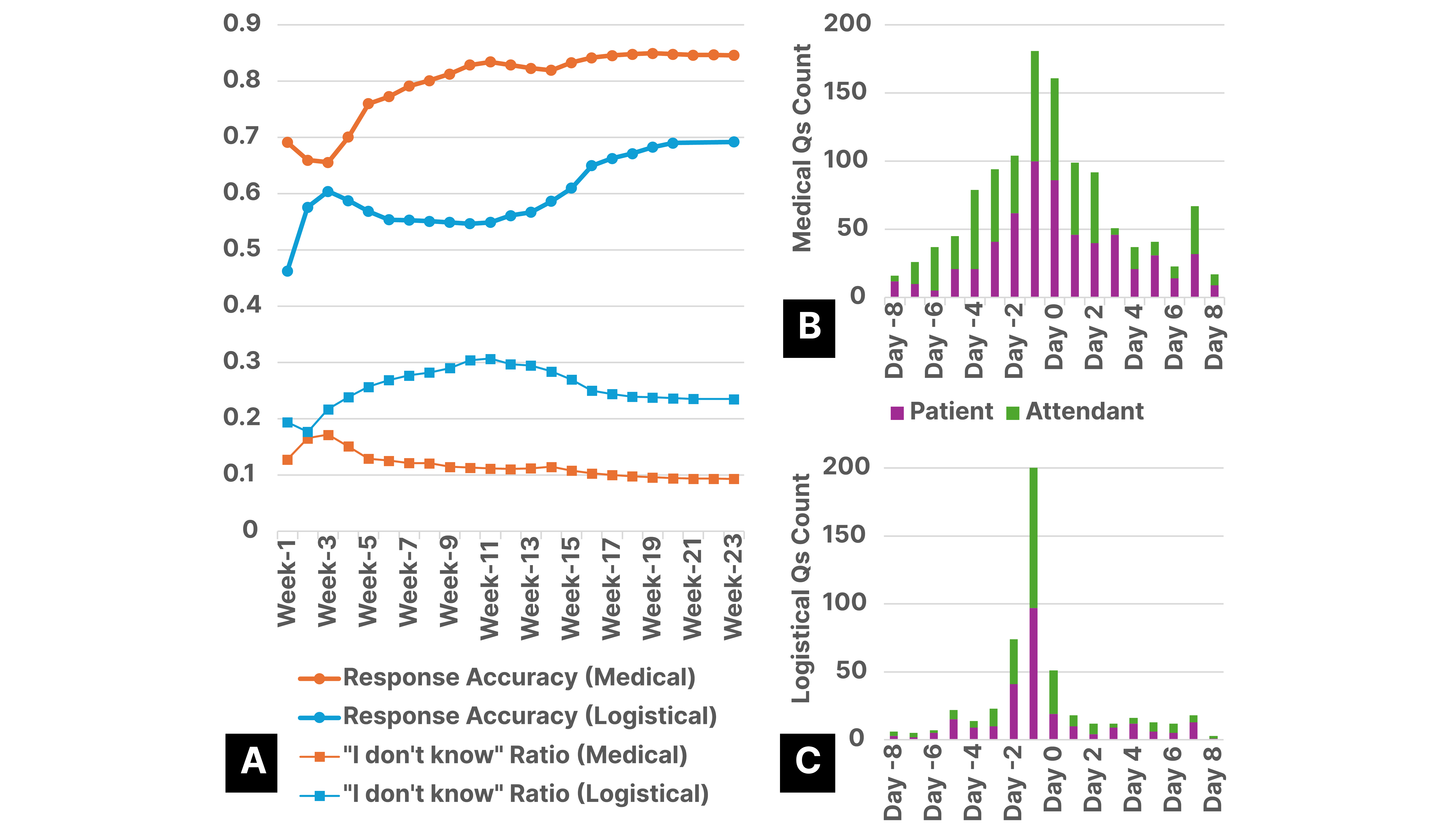}
    \caption{\textbf{(A)} Bot's performance for medical and logistical questions over the 24-week deployment, including its accuracy and completeness, based on the proportion of `\textit{Yes}' responses from experts, and the proportion of bot's ``\textit{I don't know}'' answers. In the first four weeks, 65.60\% of the bot's answers were marked as `accurate and complete', which increased to 84.62\% in the final four weeks. \textbf{(B)} Distribution of medical questions asked by patients and attendants relative to the day of surgery (Day 0). \textbf{(C)} Distribution of logistical questions asked by patients and attendants relative to the day of surgery (Day 0).}
    \label{fig:charts_botperformance_querytypedistribution}
    \Description{This consists of three figures. The left-hand Figure A is a multiline graph illustrating the accuracy and 'I don't know' ratio in the bot's responses to medical and logistical questions over time. While the accuracy for both types of questions improves over time, the accuracy for medical questions is significantly higher than for logistical questions. The 'I don't know' ratio for medical questions plateaus early, whereas for logistical questions, it initially increases and then gradually decreases. Overall, the 'I don't know' ratio is lower for medical questions than for logistical questions. The right-top Figure B is a stacked bar graph of displaying the distribution of medical queries across time for patients and attendants. There is a notable peak on Day -1 (the day before surgery), followed by the second highest number of questions on the surgery day itself. There isn't much difference between patients and attendants. The right-bottom Figure C is a stacked bar graph displaying the distribution of logistical questions across time for patients and attendants. Similar to Figure B, there is a notable peak on Day -1 (the day before surgery), followed by the second highest number of questions on the surgery day itself. The difference between the highest and second highest days is drastic. It also illustrates the decline in attendant activity after the surgery day for logistical questions.}
\end{figure*}
The top three medical questions were related to `Dos and don'ts post-surgery' (11.50\% of messages, e.g., ``\textit{Can I have some drinks? My cataract surgery is done.}''), `Medication' (7.41\%), and `Surgery preparation' (6.37\%).
Similarly, the top three logistical questions included `Surgery schedule' (11.98\%), `Hospital contact number' (5.33\%), and `Appointment scheduling' (4.36\%).
We analyzed how these themes differed between patients and attendants.
Both groups shared concerns about the surgery schedule, 
and sought clarification on general post-operative dos and don'ts
and post-surgical complications,
such as 
``\textit{How long will it take for the swelling to reduce after surgery?}''. 
However, distinct differences emerged in their information-seeking behaviours. 
Patients focused more on specifics such as food (e.g., ``\textit{Can I take coffee/tea before the surgery?}''), bathing, exercise, and screen time, comprising 23.44\% of their messages versus 10.10\% for attendants.
In contrast, attendants, acting as caregivers, inquired more about broader surgery preparation (e.g., ``\textit{If patient is diabetic, what are the precautions we need to take?}'') and medication-related queries (e.g., ``\textit{Do the eye drops have to be applied for both the eyes or just the eye which is getting operated?}'').

Other notable themes included `Anthropomorphization of bot' (11.21\%), `Fragmented questions' (1.37\%) (e.g., ``\textit{No one has called us?}'', ``\textit{For operation}'', and ``\textit{Which number to call?}''), and `Double-barrelled questions' (1.17\%), where multiple questions were asked in a single message.

\bheading{Temporal Usage Patterns}
We categorized usage into three periods relative to the day of surgery: pre-surgery (three days before surgery), on-surgery (the day of surgery), and post-surgery (three days after surgery).
A one-way RM-ANOVA showed a significant effect of these periods on the number of questions asked ($F_{2,494}$=32.39, p<0.001).
Tukey's pairwise comparison found that questions asked pre-surgery (2.73$\pm$3.23 questions/participant) were significantly more than post-surgery (1.15$\pm$2.83) and on-surgery (0.85$\pm$2.02), with p<0.001.
Specifically, the highest number of questions were asked on the day before surgery (Figure \ref{fig:charts_botperformance_querytypedistribution}B, C). 
This day accounted for 14.27\% of medical questions and 38.19\% of logistical questions respectively.
This increased pre-surgery questioning is likely due to the the heightened anxiety patients and attendants experience before surgery~\cite{ramirez2017anxietycataract}.

Medical queries were mainly asked within 7 days ($\pm$3 day) of surgery (Figure \ref{fig:charts_botperformance_querytypedistribution}B), displaying a relatively even distribution across the days.
In contrast, logistical questions were concentrated within 3 days ($\pm$1 day) of surgery, with a significant peak on the day before surgery (Figure \ref{fig:charts_botperformance_querytypedistribution}C) when patients and attendants repeatedly inquired about the next day's schedule.
Although no statistically significant correlation was found between the day of the week and the number of questions, Saturday saw the highest number of queries.
Regarding the time of day, a one-way ANOVA indicated a significant effect on the number of questions asked ($F_{3,918}$=8.95, p<0.001).
Tukey's pairwise comparison found questions asked in the evening 3pm-6pm (1.87$\pm$2.73) to be significantly more than afternoon 11am-2pm (1.22$\pm$2.25) and night 7pm-10pm (0.95$\pm$2.55), with p<0.01. 
This increase in evening activity may be attributed to the effectiveness of the 4pm reminder message sent by \cataractbot{}. 

\subsection{Expert Interaction}
\label{sec:finding-expert}
Of the 1797 bot-generated answers sent to experts for verification, 75.96\% were marked as `\textit{Yes}' indicating they were `accurate and complete', while 17.75\% were marked as `\textit{No}'.
Among the latter, 45.45\% were ``\textit{I don't know}'' responses.
On average, expert verification took 160.06$\pm$134.74 minutes.
Although responses marked as `\textit{No}' took longer to verify (162.76$\pm$141.76 minutes) compared to those marked as `\textit{Yes}' (159.47$\pm$133.21 minutes), this difference was not statistically significant.
The average time between marking a response as `No' and providing a correction was 1.59$\pm$2.65 minutes, suggesting that experts typically provided corrections soon after identifying the bot-generated answer as incorrect, which is ideal for patients and attendants.
There was also no significant time difference between verification by doctors (158.90$\pm$135.28minutes) and patient coordinators (164.02$\pm$132.98 minutes), although logistical questions were escalated more often (69.00\%) than medical ones (51.74\%).
On average, expert-corrected responses were 18.37\% longer than the bot's original responses. This difference was statistically significant with $t_{309}$=3.11, p<0.01.

Experts more often marked logistical responses as incorrect (30.54\%) compared to medical ones (15.47\%).
This was partially due to patient-specific logistical questions (19.93\%), such as ``\textit{I want to postpone my surgery.}'' and ``\textit{When can I expect the call from the insurance team?}''. 
In contrast, only 9.35\% of incorrect answers pertained to patient-specific medical questions, like ``\textit{Which lens would be used for my surgery?}.
Since the bot lacked access to patient's health records, it could not provide answers for such queries, and 
experts' responses varied accordingly.
These patient-specific corrections could not be added to the bot's knowledge base as they were not universally applicable.


Interestingly, we found logistical questions were also ignored (i.e., remained unverified by experts) more frequently (25.14\%) compared to medical ones (1.26\%).
These was mainly due to the surge of logistical questions asked on the day before surgery (78.45\%), with repeated questions contributing to the issue. 
Among the ignored questions, 12.08\% were marked as incorrect by experts, but no subsequent correction was provided despite escalations and reminders. 
This may be attributed to the experts' high workload and the effort required to formulate a correction.

\begin{figure*}
  \includegraphics[width=\textwidth]{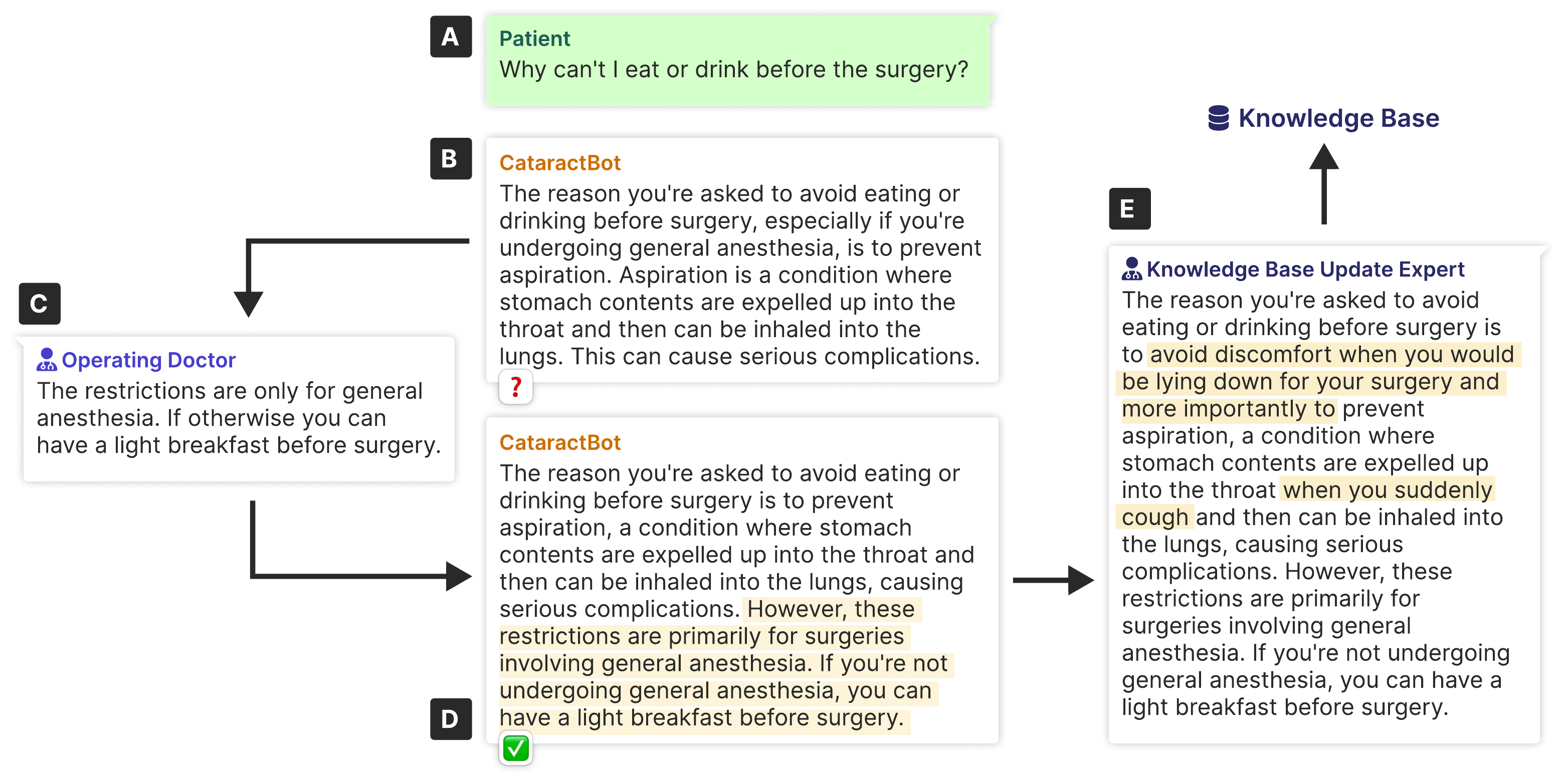}
  \caption{Life cycle of information on \cataractbot{}. \textbf{(A)} Patient's message. \textbf{(B)} Bot's initial answer. \textbf{(C)} Expert's correction. \textbf{(D)} Bot's updated answer with changes highlighted. \textbf{(E)} Knowledge Base Update Expert's edited version of the answer with changes highlighted, added to the knowledge base.}
  \Description{The bot's initial answer to the patient's question 'Why can't I eat or drink before surgery' undergoes edits when information is added by the operating expert (that the restrictions apply only if the patient is undergoing general anesthesia) and the knowledge base update expert (that food restrictions are intended to ensure patient comfort and avoid triggering aspiration through coughing). The final answer is then added to the knowledge base.}
  \label{fig:expertverifications}
\end{figure*}

\bheading{Operating Experts' Correction Analysis}
Thematic analysis of the experts' corrections (Figure \ref{fig:expertverifications}C,D) revealed nine distinct types, with the four most common described here.
First, `Adding new information' (Figure \ref{fig:expertverifications}C,D) was the most frequent correction type for both medical (65.85\%) and logistical (30.45\%) questions, wherein experts introduced new information that the bot lacked.
This category primarily addressed the bot's ``\textit{I don't know}'' responses, constituting 49.28\% of such corrections.
Second, experts performed `Factual corrections' more frequently for medical answers (7.80\%) compared to logistical answers (2.06\%). 
These medical corrections often reflected expert-specific preferences.
For example, when asked ``\textit{Can the patient eat oily food?}'', the bot suggested continuing their regular diet post-surgery, while an expert instead advised ``\textit{Avoid oily food after surgery as it can lead to cough n strain on d eyes}''.
Third, experts included `Clarifying questions' in 9.38\% of corrections.
This was primarily because 22.43\% of the bot's ``\textit{I don't know}'' responses were due to unclear or incomplete questions.
For instance, when faced with unclear questions due to translation or transcription issues, expert often requested, ``\textit{Please repeat the question, what do you mean?}'.
Similarly, in response to patient-specific questions, such as about lenses or surgery schedules,rts asked: ``\textit{Please share your MRN number [patient identification number]}''.
However, the bot was unable to handle multi-turn conversations, resulting in repeated ``\textit{I don't know}'' responses even after patients provided the requested information.
Fourth, experts responded with `Redirection' (4.24\%) to other sources of information.
For instance, coordinators advised patients to contact the insurance or inpatient desk.
In some cases, doctors recommended the patients to visit the hospital for further evaluation.

\bheading{Knowledge Base Update Expert's Edits Analysis}
Of the answers reviewed by the knowledge base update expert, 70.69\% were approved for inclusion in the knowledge base. 
Among these, 48.79\% were accepted without any modifications.
For the rest, we identified four key themes of edits.
First, the most common edit, similar to the operating expert, was `Adding new information' (78.85\%).
On average, answers added to the knowledge base (Figure \ref{fig:expertverifications}E) were 84.28\% longer than the original expert's answer (Figure \ref{fig:expertverifications}D).
Second, in 13.33\% of cases, the update expert `Generalized information' by replacing specific dates, times, or staff names with broadly applicable terms relevant to all patients.
Third, `Removing details' accounted for only 6.67\% of their edits. 
For example, when an attendant asked, ``\textit{What are some recommended activities the patient can do during recovery?}'', the operating expert replied at length, mentioning that ``\textit{You can read if comfortable with existing glasses or at review check if you can get temporary reading glasses.}''.
However, the knowledge base update expert removed the complicated qualifier, instead simply saying ``\textit{During your recovery, you can resume reading...}''.
Fourth, `Factual corrections' were made to 7.62\% of answers.
Most such corrections reflected differences in doctors' personal preferences.
For example, when an attendant asked, ``\textit{Can I stitch on sewing machine after the operation?}'', the operating doctor responded ``\textit{It's better to wait for 1 month}'', while the knowledge base update expert updated it to ``\textit{around 3 weeks}''.
Conflicting recommendations among doctors caused inconsistencies in the knowledge base.
For example, \cataractbot{} initially responded, ``\textit{You can take a head bath after 10 days post-surgery.}'' which was verified as `accurate and complete' by an operating doctor another doctor refuted the recommendation, correcting it to ``\textit{Ideally wait for 2 weeks.}''
The knowledge base update expert added this correction to the knowledge base as well.
As a result, the bot provided conflicting responses, such as ``\textit{10 days}'', ``\textit{2 weeks}'', or even ``\textit{10-14 days}'', which individual doctors continued to correct as per their preferences.
\section{Discussion}

We build upon the work of \citet{ramjee2024cataractbot} and present several novel findings. 
While they observed that attendants, being younger and tech-savvy, asked more questions than patients, we found no significant difference in the number of questions asked between the two groups.
This suggests that the bot was equally useful for all users and the WhatsApp-based system engaged older users effectively.
Further, their hypothesis that less educated, non-English speakers would prefer audio messages (in Indic languages) was not supported by our findings---audio was the least preferred modality across both patients and attendants, and we found no significant correlation between language choice and modality, or between education level and modality.
Instead, we note that more educated participants tended to ask more medical questions.
Additionally, we found the highest number of both medical and logistical questions were asked on the day before surgery, rather than on the day of surgery as \citet{ramjee2024cataractbot} reported.
This highlights the importance of pre-surgery informational support, particularly when in-person access to hospital experts is not available.
Furthermore, we contribute a qualitative analysis of edits made by operating and knowledge base update experts, revealing that most corrections involved adding new information, with repetitions and contradictions emerging as the knowledge base grew.
Finally, we provide quantitative evidence improved bot response over time, reducing expert workload, which the previous study~\cite{ramjee2024cataractbot} could only suggest anecdotally.

Below, we discuss design implications for LLM-powered experts-in-the-loop chatbots, based on insights from our large-scale, in-the-wild deployment study.

\bheading{Proactive Design to Support Predictable Information Seeking}
Our patients often sought similar types of information around the same times relative to the surgery date.
The same holds true for attendants.
Currently, as outlined in \citet{ramjee2024cataractbot}'s framework, \cataractbot{} operates reactively.
For future deployments of similar LLM-powered informational bots, we recommend incorporating a proactive information dissemination approach.
This involves the bot initiating contact with end-users via push notifications at key moments, delivering relevant information based on observed information needs across users and temporal usage patterns.
This proactive model could replace reminder messages and potentially enhance user engagement.
To support this, we propose a corpus of pre-verified information, consisting of answers that have been repeatedly approved by experts. 
In addition to notifications, this corpus could facilitate real-time, expert-approved answers and reduce the workload of experts.

\bheading{Personalization}
We found that logistical questions resulted in more ``\textit{I don't know}'' answers and inaccuracies compared to medical questions, primarily because logistical queries require patient-specific responses.
Verifying such answers proved to be a repetitive and unrewarding task for coordinators, as the information could not be added to the bot's knowledge base, leaving their future workload unchanged.
To address this issue, we propose integrating LLM-powered informational chatbots with sources of institutional knowledge. 
For \cataractbot{}, this could involve allowing it to query the hospital's patient management system to provide patient-specific responses.
However, this approach raises concerns about interoperability and privacy, as the system would potentially have access to personally identifiable user data~\cite{weng2012ehr,yang2023talk2care}.
We emphasise that users must have transparency and control over the data they share with such personalized LLM-powered bots.
As \citet{yang2023talk2care} suggests, the system should regularly remind users of the risks associated with sharing sensitive data and offer alternative ways to resolve their queries.

\bheading{Conversational Design}
In our study, we observed that conversations often failed for unclear or incomplete questions, leading the bot to respond with ``\textit{I don't know}'' and causing experts to struggle with providing answers without additional context or follow-up information.
To address this issue, the bot should be capable of identifying such questions---potentially using a Small Language Model to minimize cost and latency. 
It should then engage in multi-turn conversations with users to gather necessary clarification or enable them to build complex requests, before relaying the exchange to the expert for verification.
Additionally, \cataractbot{} currently considers only the last two queries when generating a response.
Prior work has mentioned the benefits of integrating long-term conversational history into LLM systems, enabling proactive follow-ups and more tailored recommendations~\cite{yang2023talk2care, bernstein-agent-uist}.
Incorporating a long conversation history is challenging due to token limitations in LLM inputs, hence generating conversational summaries could be explored as an alternative approach.

\bheading{Knowledge Base}
In our deployment, we observed that multiple different answers to the same or similar questions were sometimes added to the knowledge base.
We propose that those managing the knowledge base of informational bots, should have a clear overview of existing content in relation to the new question, to make an informed decision.
Additionally, previously rejected questions should not be resubmitted to the update expert.
We also noticed that these differing answers were mainly due to varied recommendations based on individual doctors' preferences.
To avoid contradictions, we suggest creating expert-specific partitions within a broader knowledge base. 
Standard subjective answers could even be parameterized, allowing experts to proactively `fill in the blanks' based on their preferences.
This decentralization of knowledge base control would reduce the burden on a single knowledge base update expert and democratize such expert-in-the-loop systems to account for the voices of individual experts.

\bheading{Translation and Transcription}
We found that transcription and translation errors, along with minimal use of audio messages, highlight the need for improvements in the services used.
For future deployments of LLM-powered bots, particularly in low-resource languages, we propose the following enhancements:
(a) Implement a dictionary of common errors across languages to serve as a look-up resource.
(b) Provide the LLM with both the original language version and its English translation of each query, for improved answer generation.
(c) Display a transcription of any audio question back to the information seeker, allowing them to verify or clarify what the bot understood.
(d) Offer experts access to the original audio of the user's question on demand, in addition to the translated text, as recommended by \citet{ramjee2024cataractbot}.
These serve as additional safeguards in cases where language technologies fail.
However, \citet{wei2022proactivesmartspeakers} caution that the advantages of using rich audio data must be balanced against privacy risks, as voice messages may unintentionally transmit personal information or identifiers.
It is essential to ensure that users are fully aware that their original audio messages will be shared with experts.

\bheading{\textit{Limitations}}
Our study has a key limitation. 
While the in-the-wild setting at \anonymousHospital{} enabled scalability, it limited our control over participant recruitment and training.
We could not ensure the accuracy of participant data, such as phone numbers,
during onboarding.
The inconsistent training methods and reliance on manual explanations from patient coordinators may have contributed to some onboarded participants not engaging with the bot, resulting in their exclusion from the study.

\section{Conclusion}

We conducted a large-scale in-the-wild 24-week deployment study of \cataractbot{}, involving 318 patients and attendants who sent 1,992 messages, receiving responses verified by seven experts.
Our analysis revealed that both patients and attendants predominantly used the bot to address their medical questions, focusing on predictable themes such as precautions and dos and don'ts post-surgery.
Their activity peaked on the day before surgery. 
\cataractbot{} rarely hallucinated, accurately classified questions and incorporated expert corrections. 
Experts rated 84.52\% of its medical answers as accurate and complete.
Their corrections mainly involving adding new information, which reduced \cataractbot{}'s ``\textit{I don't know}'' answers by 7.84\% and improved its accuracy and completeness by 19.02\% over the study period.
We note, however, that experts often ignored patient-specific questions due to a lack of incentives, as these non-generalized inputs would not be added to the knowledge base to reduce their future workload.
To address this, we emphasize the importance of integrating such bots with institutional sources of user-specific information.
Finally, as the knowledge base expanded, we observed repeated information and conflicting recommendations from different doctors.
We propose creating expert-specific partitions within the knowledge base. 
We hope these insights will inform the design and development of LLM-powered expert-in-loop chatbots for domains including and beyond healthcare.

\begin{acks}
Many thanks to the staff of \anonymousHospital{} for their time and patience.
\end{acks}

\bibliographystyle{ACM-Reference-Format}
\bibliography{main}



\end{document}